\documentclass[12pt,twoside]{article}
\usepackage{xrb2007}
\usepackage{graphics}

\begin{document}

\session{Jets}

\shortauthor{Soleri et al.}
\shorttitle{Multiwavelength Observations of Swift~J1753.5-0127}

\title{Multiwavelength Observations of the Black Hole Candidate Swift J1753.5-0127}
\author{P. Soleri, D. Altamirano}
\affil{Astronomical Institute "A. Pannekoek", University of Amsterdam, Kruislaan 403, 1098 SJ
Amsterdam, the Netherlands}
\author{R. Fender}
\affil{School of Physics and Astronomy, University of Southampton, Highfield, Southampton SO17 1BJ, UK}
\author{P. Casella, V. Tudose, D. Maitra, R. Wijnands}
\affil{Astronomical Institute "A. Pannekoek", University of Amsterdam, Kruislaan 403, 1098 SJ
Amsterdam, the Netherlands}
\author{T. Belloni}
\affil{INAF - Osservatorio Astronomico di Brera, via Emilio Bianchi 46, 23807 Merate (LC), Italy}
\author{J. Miller-Jones}
\affil{NRAO, 520 Edgemont Road, Charlottesville, VA 22903, USA}
\author{M. Klein-Wolt and M. van der Klis}
\affil{Astronomical Institute "A. Pannekoek", University of Amsterdam, Kruislaan 403, 1098 SJ
Amsterdam, the Netherlands}

\begin{abstract}
We present preliminary results from the analysis of simultaneous multiwavelength observations of the
black hole candidate Swift J1753.5-0127. The source is still continuing its outburst started in May 2005,
never leaving the Low/Hard State. In the X-ray energy spectra we confirm evidence for a thermal component 
at a very low luminosity possibly extending close to but not at the innermost stable orbit. This is
unusual for black hole candidates in the Low/Hard State. Furthermore, we confirm that its radio emission
is significantly fainter than expected from the relation observed in other Black Hole Candidates between
the observed radio/X-ray fluxes.
\end{abstract}

\vspace{-0.9cm}
\section{Introduction}   \label{par:intro}
Swift J1753.5-0127 was discovered with Swift/BAT on
2005 May 30 (Palmer et al. 2005). Spectral
and timing analysis performed with Swift/XRT and RXTE/PCA (Morris et
al.\ 2005, Morgan et al.\ 2005) revealed a hard power-law spectrum and
a 0.6 Hz QPO that might indicate the presence of a black hole in the system.
The source was also detected in the radio band with MERLIN at a
flux density of 2.1 mJy at 1.7 GHz (Fender et al.\ 2005), probably
indicating jet activity.

\begin{figure}
\begin{tabular}{c}
\resizebox{16cm}{!}{\includegraphics{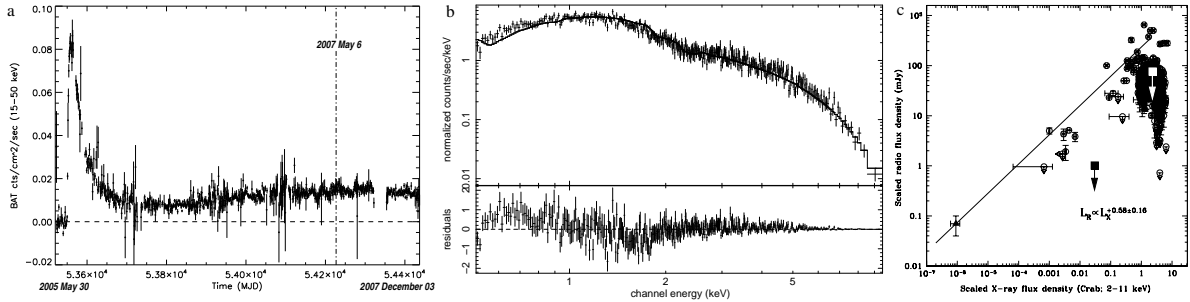}}
\end{tabular}
\vspace{-0.3cm}
\caption{{\it Panel a:} Swift/BAT light curve of Swift J1753.5-0127. The vertical dashed-dotted line indicates
the beginning of our multiwavelength
observational campaign. {\it Panel b:} Swift/XRT energy spectrum for 2007 July 03, 1.7 ks exposure. The
spectrum is fitted with an absorbed power law. From the plot of the residuals one can see that a thermal
component is needed to get an acceptable fit. {\it Panel c:} figure 4 from Gallo et al.
2006 where we added the Swift J1753.5-0127 data point (black square with black thick arrow for a distance of 1 kpc, white
square with white thick arrow for a distance of 8.5 kpc) taken from a simultaneous Swift/XRT - WSRT observation
on 2007 July 08.}
\label{fig}
\end{figure}

After reaching a flux peak of 120 mCrab on 2005 July 1 (Morgan et al.\ 2005),
the source flux began to decay, but has stalled for months in its decline at
a level of $\sim$20 mCrab (2-20 keV). This is an unusual behaviour
for a black hole transient, but even more unusual is the slow rebrightening, which occurred
between 2006 June and 2007 July-August, as can be seen in the Swift/BAT light curve in Figure \ref{fig}
(panel a). At present, the X-ray light curve of Swift J1753.5-0127
suggests a slow decay. The source never left the Low/Hard State (LS) during the whole outburst
(Cadolle Bel et al.\ 2007, Zhang et al.\ 2007).

Miller et al. 2006a, analysing XMM-Newton observations collected during
2006 March, estimated a disk component extending to the innermost stable
orbit at \\
$L_{\rm X}/L_{{\rm Edd}} \simeq 0.003\, (d/8.5\, {\rm kpc})^2\,
(M/10M_{\odot})$. This is one of only three black hole candidates (BHCs) that show a thermal component in LS
at a very low X-ray luminosity, since they are usually found not to have such a component (Esin et al. 1997,
Fender, Belloni \& Gallo 2004; but see exceptions: Miller et al. 2006b for GX 339-4; Rykoff et al. 2007
for XTE J1817-330).

Cadolle-Bel et al. 2007 analysing radio VLA observations, reported an average flux of 0.7 $\pm$ 0.1
mJy at 8.5 GHz, when the unabsorbed X-ray flux was  $1.5 \times 10^{-9} {\rm erg}/{\rm cm}^2/{\rm sec}$ (2-11 keV).
With these values the source is extremely radio fainter than expected from the empirical correlation found
by Gallo et al. 2006 between the radio and the X-ray luminosity (scaled to a distance of 1 kpc)
of a number of BHCs in LS and in quiescence: $L_{\rm R} \propto L_{\rm X}^{0.58\pm0.16}$. 

\vspace{-0.4cm}
\section{Observations}
We are currently observing the source with several instruments in different bands of the electromagnetic
spectrum.

{\bf X-ray - RXTE$+$Swift:} we started monitoring the source weekly on 2007 May 6th
with the {\it Rossi X-ray Timing Explorer} satellite ($\sim$3-100 keV, PCA+HEXTE) and we are
currently observing it. The vertical dashed-dotted line in Figure \ref{fig} (panel a) indicates the beginning of the
RXTE observing campaign.
Swift/XRT ($\sim$0.5-10 keV) pointed at the source 9 times between 2007 May 28th and 2007
July 15th, with a typical exposure of $\sim$2 ksec. 

{\bf Radio - WSRT:} we observed the source in radio 4 times with the Westerbork Synthesis Radio Telescope
on 2007 July 1st, 2007 July 8th, 2007 July 15th and 2007 July 22nd, at 5 and 8 GHz. RXTE and Swift observed
the source simultaneously during the first three radio pointings, only RXTE observed the last one.

{\bf Optical/IR - SMARTS:} the SMARTS telescope observed the source 47 times between 2007 July 9th and 2007
September 30th, in the I, V and H band. The usual exposures were 15 min, enough to get a typical signal to noise ratio
higher than 50.      

\vspace{-0.4cm}
\section{Preliminary results and conclusions}
Here we present preliminary results from our multiwavelength analysis, focusing on the two issues pointed out at
the end of section \S \ref{par:intro}

Figure \ref{fig} (panel b) shows a Swift/XRT energy spectrum taken on 2007 July 03. Fitting the spectrum with an
absorbed power law we obtain a fit with a reduced chi squared  $\chi_{\nu}^{2}$ = 1.55 (fixing the galactic
absorption $N_H$ to the value found by Miller et al. 2006a, $N_H=2.3\times 10^{21}$ cm$^{-2}$).
The residuals can be significantly reduced by the addition of a thermal component
with $kT_{bb} = 0.20\pm0.02$ keV (significance 4.2 $\sigma$): the new fit gives a chi squared
$\chi_{\nu}^{2}$ = 1.23. This three-component model gives an
absorbed flux of $7.4 \times 10^{-10} {\rm erg}/{\rm cm}^2/{\rm sec}$ (0.5-10 keV): this corresponds to $L/L_X = 0.005
(d/8.5 {\rm kpc})^2 (M/10M_{\odot})$. The normalization gives an inner radius $R_{in}$ = 72$^{+17}_{-9} (d/8.5 {\rm
kpc})
/ cos\,i$ km. 
In all our spectral fits (but one that gave not consistent results) we found a temperature for the thermal
component consistent with $kT_{bb} =
0.20$ keV and an inner disc radius $R_{in}$ in the range ($46^{+21}_{-8} < R < 72^{+17}_{-9}$) km. Spectra have been
fitted fixing the Galactic absorption $N_H$ to the value found by Miller et al. 2006a.

Figure \ref{fig} (panel c) shows simultaneous X-ray/radio flux densities for a number of BHCs
(scaled to a distance of 1 kpc) used by Gallo et al. 2006 to find the empirical correlation
between the radio and the X-ray luminosity mentioned in \S \ref{par:intro} We measured
Swift J1753.5-0127 X-ray and radio flux on 2007 July 08: in radio, with WSRT, we did not detect the
source but we could give a 3 $\sigma$ upper limit on its flux of 1.1 mJy/beam. With Swift/XRT we measured a
X-ray flux (2-11 keV) of $6.2 \times 10^{-10} {\rm erg}/{\rm cm}^2/{\rm sec}$ (26 mCrab). We added two points to the plot, a black
and a white square, for a source distance of 1 kpc and 8.5 kpc respectively.

Our preliminary results from simultaneous multiwavelength observations of the BHC Swift J1753.5-0127 confirm
previous results:

{\it (i)} in all our spectral fits (but one) we found results partially consistent with Miller et al. 2006a :
a thermal component with a temperature consistent with the one found by Miller et al. 2006a ($kT_{bb} =
0.22\pm0.01$ keV) is needed. The inner radius of the disc blackbody associated with the thermal component is
higher than the radius found by Miller et al. 2006a ($R_{in}$ = $30\pm5$ km) and not consistent with it (only
in one case it is consistent within the errors). A broad
band spectral analysis is clearly needed to model Swift J1753.5-0127 emission, putting light on the origin of
the thermal emission.

{\it (ii)} Figure \ref{fig} clearly shows that the source is still less luminous in radio than
expected from the $L_{\rm X} - L_{\rm R}$ correlation, as previously reported by Cadolle Bel et al. 2007.
This might weaken the identification of Swift J1753.5-0127 as a BHC, not supported by any dynamical
measurement of the mass of the compact object: NSs are usually much radio fainter than BHCs at the same X-ray luminosity
(Fender \& Hendry 2000). However note that other LS sources are also ``partially radio quenched''
(e.g. XTE J1720-318, IGR J17497-2821, XTE J1650-500, 1E 1740.7-2942, GRS 1758-258; Gallo 2007). 
This issue will need further studies and observations.

\vspace{-0.45cm}

\end{document}